\documentclass[fleqn,twoside]{article}
\usepackage{espcrc2}

\usepackage{graphicx}
\usepackage[figuresright]{rotating}

\newcommand{\AmS}{{\protect\the\textfont2
  A\kern-.1667em\lower.5ex\hbox{M}\kern-.125emS}}
\newcommand{\beq}{\begin{equation}}
\newcommand{\eeq}{\end{equation}}
\newcommand{\beqa}{\begin{eqnarray}}
\newcommand{\eeqa}{\end{eqnarray}}

\def \branch{{\cal B}}

\def\be{\begin{equation}}
\def\ee{\end{equation}}

\def\bea{\begin{eqnarray}}
\def\eea{\end{eqnarray}}
\def\be{\begin{equation}}
\def\ee{\end{equation}}

\def\beq{\begin{equation}}
\def\eeq{\end{equation}}

\def\qbar{\overline q}

\def\q5q{\qbar{{\lambda_a}\over 2} i\gamma_5 q}

\def\to{\rightarrow}

\def \eff{\hbox{eff}}  
\def \nn{\nonumber}

\def \gh{\vphantom{$\fbox{\Big[}$}}
    
\def \cl#1{{#1\%\ \mathrm{C.L.}}}
\begin{document}

\title {\vspace*{-1.0cm}
\begin{flushright}
{\small CERN-TH/2002-284\\
October 2002}\\
\end{flushright}
Next-to-leading Order Calculations of the Radiative and 
Semileptonic Rare $B$ Decays in the Standard Model and Comparison with 
Data}

\author{A. Ali\address[MCSD]{Deutsches Elektronen-Synchrotron DESY,\\ 
       Notkestra\ss e 85, D-22603 Hamburg, Federal Republic of Germany   
}%
\thanks{Present address: CERN (-TH-), CH-1211 Geneva 23.}}
       

\begin{abstract}
We review some selected rare $B$ decays
calculated in next-to-leading order accuracy in the Standard Model (SM). 
These include the radiative decays $B \to (X_s,K^*,\rho )\gamma$ 
and the semileptonic decays $B \to (X_s,K,K^*) \ell^+ \ell^-$, for 
which new data from the BABAR and BELLE collaborations have been presented 
at this conference. SM is in agreement with the 
current measurements within the theoretical and experimental errors. The 
impact of rare $B$-decays on the CKM  phenomenology is quantitatively 
discussed. 
\vspace{1pc}
\end{abstract}

\maketitle

\section{INTRODUCTION AND OVERVIEW}
First measurements of rare $B$ decays
$B \to K^* \gamma$ \cite{Ammar:1993sh} and
$B \to X_s\gamma$~\cite{Alam:1995aw} were reported by CLEO in 1993 and 
1995, respectively..
 Since then, a wealth of data on
these decays has become available through the subsequent work of the CLEO
collaboration \cite{Chen:2001fj,Coan:1999kh}, the ALEPH collaboration at 
LEP \cite{Barate:1998vz}, and more recently from the
experiments at the B factories, BABAR~\cite{Jessop02} and 
BELLE~\cite{Nishida02}, which now dominate this field.
We quantify the impact of $B \to X_s \gamma$ measurement on
the CKM phenomenology. 

Concerning the exclusive decays $B \to 
K^* \gamma$, the NLO calculations of the decay widths were 
completed last year \cite{bdgAP,Bosch:2001gv,Beneke:2001at}. Radiative 
decays provide a cleaner test of the underlying theoretical framework
of perturbative QCD factorization which has been invoked initially to
calculate the two-body exclusive non-leptonic $B$ decays,
such as $B \to \pi \pi$ \cite{Beneke:1999br}.
Quantitative {\it rapport} of theory and experiment in these decays is 
on hold at present due to the imprecise knowledge of the form factor, as  
discussed later in this talk.
      
A big leap forward has been reported in the 
experimental study of the Cabibbo-suppressed
decays $B \to \rho \gamma$ and $B \to \omega \gamma$ at this conference
\cite{Jessop02,Nishida02}. These decays which enact $b \to d \gamma$ 
transition at the quark level yield complementary information 
on the CKM parameters and hence are potentially very important. We work 
out the constraints following from the current upper limits on
the $B \to \rho \gamma$ branching ratios \cite{Jessop02,Nishida02}.
Apart from these, the 
Cabibbo-suppressed radiative decays $B \to \rho \gamma$ allow us to study
isospin- and CP-asymmetries, which in the SM have a sensitive dependence 
on the angle $\alpha$ of the unitarity triangle 
\cite{Ali:2000zu,bdgAP,Bosch:2001gv}. We present the SM expectations 
for the decay rates and asymmetries, which  are 
also potentially of interest  in searching for physics beyond-the-SM 
\cite{Ali:2000zu,Ali:2002kw}.

The semileptonic decays $B \to (X_s, K,K^*) \ell^+ \ell^-$ ($\ell^\pm 
=e^\pm, \mu^\pm$) are the new additions to the flavour changing 
neutral current FCNC decays, where B factories have made first
experimental inroads \cite{Nishida02,Richman02}. Of particular interest is 
the inclusive decay
$B \to X_s \ell^+ \ell^-$, whose measurement has been reported at this 
conference by the BELLE collaboration \cite{Nishida02}.
These decays have been calculated in the SM to the same degree of 
theoretical precision as their radiative counterpart $B \to X_s \gamma$,
and take into account the explicit $O(\alpha_S)$ corrections
to the dilepton invariant mass distribution 
\cite{Bobeth:1999mk,Asatrian:2001de}, the forward-backward asymmetry of
the charged lepton \cite{Ghinculov:2002pe,Asatrian:2002va}  
and the leading power corrections in $1/m_b$ \cite{Ali:1996bm,Falk:1993dh} 
and $1/m_c$ \cite{Buchalla:1997ky}. Precise measurements of the  
dilepton invariant mass spectrum and the forward-backward asymmetry
would allow to extract the (effective) Wilson coefficients, providing 
precision tests of the SM and enabling  searches for physics beyond-the-SM 
\cite{Ali:1994bf}.

What concerns the exclusive decays $B \to (K,K^*) \ell^+ \ell^-$, 
important theoretical progress has been made recently
\cite{Beneke:2000wa,Beneke:2001at}, using ideas  
based on the large energy expansion \cite{Charles:1998dr}. In particular,
both the dilepton invariant mass spectrum and the forward-backward 
asymmetry in $B \to K^* \ell^+ \ell^-$ have been calculated in the lower 
region of the dilepton invariant mass ($s < m_{J/\psi}^2)$ to the same 
accuracy as is the case for $B \to X_s \ell^+ \ell^-$, albeit using 
perturbative QCD factorization. 
This framework has also been used to carry out a helicity analysis of 
the decays $B \to K^* \ell^+ \ell^-$ and $B \to \rho \ell \nu_\ell$  
\cite{Ali:2002qc}.
Data on $B \to (X_s,K^*,K) \ell^+ \ell^-$ are compared with
the SM estimates \cite{Ali:1999mm,Ali:2002jg} and the two are 
found to be compatible with each other.

\section{Inclusive Decay $B \to X_s \gamma$}
The theoretical framework for analyzing the transition $B \to X_s \gamma$ 
is provided by the effective Hamiltonian
\be \label{Heff}
{\cal H_{\rm eff}} =  -\frac{4 G_F}{\sqrt{2}} V_{ts}^* V_{tb}
 \sum_{i=1}^8 C_i(\mu) O_i,
\ee
where $G_F$ is the Fermi coupling constant and $V_{ij}$ are the CKM matrix 
elements. The effective Hamiltonian 
is obtained from the SM by integrating out all the particles that are
much heavier than the $b$-quark. The Wilson coefficients $C_i(\mu)$
play the role of effective coupling constants for the interaction terms
$O_i$. The generic structure of the operators $O_i$ can be seen, for 
example, in Ref.~\cite{Ali:2002jg}.

Perturbative calculations are used to determine the Wilson coefficients in
a given renormalization scheme, such as the
$\overline{\rm MS}$ scheme, at a renormalization scale, typically $\mu_b 
\sim O(m_b)$,
\be \label{cmb}
C_i(\mu_b) = C_i^{(0)}(\mu_b)+ \frac{\alpha_s(\mu_b)}{4\pi} 
C_i^{(1)}(\mu_b) + \ldots.
\ee
Here, $C_i^{(n)}(\mu_b)$ depend on $\alpha_s$ only via the ratio $\eta
\equiv \alpha_s(\mu_0)/\alpha_s(\mu_b)$, where $\mu_0 \sim m_W$. In
the leading order (LO) calculations, everything but $C_i^{(0)}(\mu_b)$
is neglected in Eq.~(\ref{cmb}). In NLO, one takes in 
addition $C_i^{(1)}(\mu_b)$ into account.

It has become customary to quote the theoretical branching ratios for
$B \to X_s \gamma$ decay with a cut on the photon energy $E_\gamma$.
In the $\overline{\rm MS}$ scheme, the results in the SM (in units of 
$10^{-4}$) are 
\cite{Gambino:2001ew,Buras:2002tp}:
\bea \label{thSM_1.6}
{\cal B}(\bar{B} \to X_s \gamma~;E_\gamma > 1.6\;{\rm GeV})
&=& (3.57 \pm 0.30), \nonumber\\
{\cal B}(\bar{B} \to X_s \gamma~; E_\gamma > 
 ~{\textstyle\frac{1}{20}}m_b)
&=& (3.70 \pm 0.31). \nonumber\\ \label{thSM_mb20}
\eea
The present world averages (also in units of $10^{-4}$) 
\bea
{\cal B}(\bar{B} \to X_s \gamma~;E_\gamma > 1.6\;{\rm GeV}) &=&
\left( 3.28 \; {}^{+0.41}_{-0.36} \right), \label{av_1.6}
\nonumber \\
{\cal B}(\bar{B} \to X_s \gamma~;E_\gamma > 
~{\textstyle\frac{1}{20}}m_b)
 &=&
\left( 3.40 \; {}^{+0.42}_{-0.37} \right)\nonumber\\ \label{av_mb20}
\eea
result from the following four measurements: BABAR 
\cite{Jessop02,Aubert:2002pd},
CLEO \cite{Chen:2001fj}, BELLE \cite{Abe:2001hk}, and ALEPH 
\cite{Barate:1998vz}. They are in good agreement with the NLO results in
the $\overline{\rm MS}$ scheme. However, there is a residual theoretical 
uncertainty on the SM estimates related to the scheme-dependence of the 
quark masses. This can be judged from the theoretical branching ratios in 
the $\overline{\rm MS}$ scheme for which one gets
${\cal B}(\bar{B} \to X_s \gamma)=(3.73 \pm 0.31) \times 10^{-4}$ 
\cite{Gambino:2001ew}, and in the pole quark mass scheme yielding
${\cal B}(\bar{B} \to X_s \gamma)=(3.35 \pm 0.30) \times 10^{-4}$
\cite{Chetyrkin:1996vx}.
Reducing this theoretical uncertainty, which currently represents an 
 error of $O(10\%)$,  requires calculation of the
next-to-next-to-leading order contribution to the decay width (i.e., 
explicit $O(\alpha_s)^2$ improvements), which is a formidable but 
doable task in a concerted theoretical effort \cite{Misiak-02}. 
Anticipated 
experimental precision at the $B$ factories will make it mandatory to 
undertake this heroic effort. 

The transition $b \to s 
\gamma$ is not expected to yield useful information on
the CKM-Wolfenstein parameters $\bar \rho$ and $\bar \eta$, which define 
the apex of the
unitarity triangle of current interest.  The test of CKM-unitarity for the 
$b
\to s$ transitions in rare $B$-decays lies in checking the relation 
$\lambda_t \simeq -\lambda_c$, which holds up to corrections of order 
$\lambda^2$. This is implicit in the theoretical decay rates quoted 
above.
 Alternatively, one can drop the explicit use of the CKM unitarity and
calculate the various contributions in the $b \to s \gamma$ amplitude from 
the current knowledge of $\lambda_c  
\simeq \vert V_{cb} \vert =(41.0 \pm 2.1) \times 10^{-3}$ and 
$\lambda_u$ from PDG \cite{Hagiwara:pw}, which
then yields $\lambda_t \simeq V_{ts} = -(47\pm 8)\times 10^{-3}$ 
\cite{AM-02}. This is
consistent with $\lambda_t=-\lambda_c$, but less precise. This 
is due to the strong mixing of the operator $O_2$ with 
$O_7$ under QCD renormalization, which enhances the 
coefficient of the $\lambda_c$ term in the $b \to s 
\gamma$ amplitude in comparison with the coefficient of the $\lambda_t$ 
term, thereby reducing the precision on $\lambda_t$.  

\section{EXCLUSIVE DECAYS $B \to (K^*,\rho) \gamma$ }

To compute the branching ratios for $B \to V \gamma$ ($V=K^*,\rho)$ 
reliably, one needs to calculate the explicit $O (\alpha_s)$ improvements
to the lowest order decay widths. This
requires the calculation of the renormalization group effects in the
appropriate Wilson  coefficients in the effective
Hamiltonian~\cite{Chetyrkin:1996vx}, an explicit $O(\alpha_s)$ calculation
of the matrix elements involving the hard vertex
corrections~\cite{Soares:1991te,Greub:1994tb,Greub:1996tg}, annihilation 
contributions~\cite{Ali:1995uy,Khodjamirian:1995uc,Grinstein:2000pc},
which are more important in the decays $B \to \rho \gamma$ but have also 
been worked out for the $B \to K^* \gamma$ decays \cite{Kagan:2001zk},
and the so-called hard-spectator contributions involving (virtual) hard
gluon radiative corrections off the spectator quarks in the~$B$-, $K^*$-,  
and~$\rho$-mesons~\cite{bdgAP,Bosch:2001gv,Beneke:2001at}.
 We discuss the decays $B \to K^* \gamma$ and $B 
\to \rho \gamma$ in turn.

\subsection{ $B \to K^* \gamma$ Decays}
The  branching ratio for the decays $B \to K^* \gamma$ can be 
written as \cite{bdgAP,Bosch:2001gv,Beneke:2001at}:
\begin{eqnarray}
{\cal B} (B \to K^{*} \gamma) & = &
\tau_B \,\frac{G_F^2 \alpha |V_{tb} V_{ts}^*|^2}{32 \pi^4} \,
m_{b, {\rm pole}}^2 \, M_B^3 \,
\nonumber \\
 & & \hspace*{-3.0cm} \times 
\left [ \xi_\perp^{(K^*)}(0) \right ]^2
\left ( 1 - \frac{m_{K^*}^2}{M^2} \right )^3
\left | C^{(0){\rm eff}}_7 +  A^{(1)}(\mu) \right |^2~.
\nonumber
\end{eqnarray}
Here, $M_B$ is the $B$-meson mass and $\xi_\perp^{(K^*)}(0)$ is the form 
factor in the large energy effective theory. It differs from the 
corresponding form factor in 
QCD, $T_1^{K^*}(0)$, by $O(\alpha_s)$ terms
calculated in Ref.~\cite{Beneke:2000wa}. Numerically, this relation can
be expressed as $T_1^{K^*}(0) \simeq 1.08 \xi_\perp^{(K^*)}$. 
The effect of the $O(\alpha_s)$ corrections is encoded in the 
function $A^{(1)}(\mu)$ and its numerical value can be expressed in terms 
of a $K$-factor \cite{bdgAP,Bosch:2001gv,Beneke:2001at}:
\begin{displaymath}
K =\frac{\left | C_7^{(0) {\rm eff}} + A^{(1)} (\mu) \right |^2}
        {\left | C_7^{(0) {\rm eff}} \right |^2},
\quad
{\rm with}
\quad
1.5 \le K \le 1.7~.
\end{displaymath}
This yields a charge-conjugate averaged branching ratio 
\begin{eqnarray}
\langle{\cal B} (B \to K^* \gamma)\rangle \simeq
(7.2 \pm 1.1)\times 10^{-5} \, \nonumber\\
& & \hspace*{-4.0cm}\times 
\left ( \frac{m_{b,{\rm pole}}}{4.65~{\rm GeV}} \right )^2
\left ( \frac{\xi_\perp^{(K^*)}(0)}{0.35} \right )^2~,
\nonumber 
\end{eqnarray}
where the default values of the most sensitive parameters are indicated.
The value given for  $\xi_\perp^{(K^*)}(0)$ is based on using the
light-cone QCD sum rules. Varying the parameters within reasonable ranges
(see, for example, \cite{bdgAP}) and adding the errors in quadrature, one 
gets
\begin{equation}
\langle{\cal B} (B \to K^* \gamma)\rangle \simeq
(7.2 \pm 2.7)\times 10^{-5}~,
\label{smbkstar}
\end{equation} 
to be 
compared with the corresponding experimental measurement 
\cite{Coan:1999kh,Jessop02,Nishida02}:
\begin{equation}
\langle{\cal B} (B \to K^* \gamma)\rangle \simeq
(4.22 \pm 0.28)\times 10^{-5}~.
\label{exbkstar}
\end{equation}
Given the large theoretical errors, there is agreement between theory and 
experiment. Since there exists 
good agreement between the SM and experiment in the inclusive decay 
$B \to X_s \gamma$, one could use the measured branching ratios for
$B \to K^* \gamma$ to determine the form factor $T_1^{K^*}(0)$, which in 
the stated theoretical context yields
$T_1^{K^*}(0) = 0.27 \pm 0.04$. This is to be compared with the light-cone
QCD sum rule result $T_1^{K^*}(0) = 0.38 \pm 0.05$ 
\cite{Ball:1998kk,Ali:1999mm},
and with $T_1^{K^*}(q^2=0) = 0.32 ^{+0.04}_{-0.02}$  
\cite{DelDebbio:1997kr}, obtained by using the Lattice UK-QCD 
simulations combined with a LC-QCD-sum-rule inspired input for 
extrapolation 
from $q^2 \gg 0$, where the Lattice simulations are actually done, to
the physical value $q^2=0$. A 
preliminary result on $T_1^{K^*}(0)$
using Lattice-QCD has also been reported at this conference 
\cite{Becirevic02}. Due to the spread in the theoretical estimates it
is too early to draw a quantitative conclusion on the
precise value of the form factor, and hence on the  perturbative QCD 
factorization method underlying the theoretical estimates in $B \to K^* 
\gamma$.

\subsection{ $B \to \rho \gamma$ Decays}
The effective Hamiltonian for the radiative decays $B \to (\rho,\omega) 
\gamma$ can be seen for the SM in \cite{bdgAP}. These
transitions involve the CKM matrix elements in the first and third
column, with the unitarity constraints taking the form $\sum_{u,c,t}
\xi_i =0$, with $\xi_i=V_{ib}V_{id}^*$. All three matrix elements  
are of order $\lambda^3$, with
$\xi_u \simeq A\lambda^3 (\bar \rho - i \bar \eta)$,
$\;\xi_c \simeq -A\lambda^3$,
and $\xi_t \simeq A\lambda^3(1-\bar \rho - i \bar \eta)$.
This equation leads to the same unitarity triangle as studied through
the constraints $V_{ub}/V_{cb}$, $\Delta M_{B_d}$ (or $\Delta
M_{B_d}/\Delta M_{B_s}$).

We shall concentrate here on $B \to \rho \gamma$ decays. As the absolute 
values of the form factors in this decay and in $B \to K^* \gamma$ decays
discussed earlier are quite uncertain, it is advisable to calculate 
instead the ratios of the branching ratios
\begin{equation}
R^\pm(\rho \gamma/K^*\gamma) \equiv \frac{{\cal B} (B^\pm \to \rho^\pm 
\gamma)}{{\cal B} (B^\pm \to K^{*\pm} \gamma)},
\label{rpm}
\end{equation}
\begin{equation}
 R^0(\rho \gamma/K^*\gamma) \equiv \frac{{\cal B} (B^0 \to 
\rho^0 \gamma)}{{\cal B} (B^0 \to K^{*0} \gamma)}.
\label{rzero}
\end{equation}
They have been calculated in the NLO accuracy and the result can be
expressed as \cite{bdgAP}:
\bea
R^\pm (\rho \gamma/K^* \gamma) &=&  \left| V_{td} \over V_{ts} \right|^2
 {\rm [PS]}~\zeta^2
(1 + \Delta R^\pm) \; , \nonumber\\
R^0 (\rho \gamma/K^* \gamma) &=& {1\over 2} \left| V_{td} 
 \over V_{ts} \right|^2
 {\rm [PS]}~\zeta^2
(1 + \Delta R^0) \; , \nonumber
\label{rapp}
\eea
where ${\rm [PS]} =(M_B^2 - M_\rho^2)^3 /(M_B^2 - 
M_{K^*}^2)^3$  represents a kinematic factor, which is 1 to a good 
approximation, and 
$\zeta=\xi_{\perp}^{\rho}(0)/\xi_{\perp}^{K^*}(0)$, with
$\xi_{\perp}^{\rho}(0) (\xi_{\perp}^{K^*}(0))$ being the form factors  
(at $q^2=0$) in the effective heavy quark theory for the decays $B \to
\rho (K^*)\gamma$. Noting that in the SU(3) limit one has $\zeta=1$,
we take $\zeta=0.76 \pm 0.10$ - a range which straddles various 
theoretical 
estimates of this quantity
\cite{Ali:1995uy,Ball:1998kk,Narison:1994kr,Melikhov:2000yu}. 
The functions $\Delta R^\pm$ and $\Delta R^0$, appearing on the r.h.s. 
of the above equations and whose parametric dependence is 
suppressed here for ease of writing, encode
both the $O(\alpha_s)$ and annihilation contributions. They have been
evaluated as \cite{bdgAP,Ali:2002kw}: 
$\Delta R^\pm =0.055 \pm 0.130$ and $\Delta R^0=0.015 \pm 0.110$, where 
the errors also include the effect of varying the angle $\alpha$ in the
currently allowed region.

There are two more observables of interest, namely the isospin-violating  
ratio  $\Delta (\rho\gamma)$ defined as 
\begin{equation}
\Delta (\rho\gamma) \equiv {\Gamma^\pm (B \to \rho \gamma) \over 2 \; 
\Gamma^0 (B  \to \rho 
\gamma)} -1 ~,
\label{isoasy}
\end{equation}
and the CP asymmetry in the decay rates, which for $B^\pm \to \rho^\pm 
\gamma$ decays is defined as
\begin{equation}
A^\pm_{CP} (\rho\gamma) \equiv \frac{\branch (B^- \to \rho^- \gamma) -
\branch (B^+ \to \rho^+ \gamma) }{\branch (B^- \to \rho^- \gamma) + 
\branch (B^+ \to \rho^+ \gamma) }.
\label{dircp}
\end{equation}
The expected values of the observables in the $B \to \rho \gamma$ system
defined in Eqs.~(\ref{rpm}) - (\ref{dircp}) and 
$A_{CP}^0(\rho\gamma)$,
the CP asymmetry in the neutral modes, 
have been recently updated in Ref.~\cite{Ali:2002kw}. Including the 
errors on the current determination of the CKM parameters, the results
are:
\bea
R^\pm (\rho \gamma /K^* \gamma) &=& 0.023 \pm 0.012 \; ,\nonumber\\
R^0 (\rho \gamma /K^* \gamma) &=& 0.011\pm 0.006 \; ,\nonumber\\
\Delta (\rho \gamma ) &=& 0.04^{+0.14}_{-0.07} \; ,\nonumber\\
A_{CP}^\pm (\rho\gamma) &=& 0.10^{+0.03}_{-0.02} \; , \nonumber\\
A_{CP}^0 (\rho\gamma) &=& 0.06 \pm 0.02 \; .
\eea  

At this conference, the BABAR collaboration has reported a significant
improvement on the upper limits of the branching ratios for the
decays $B^0(\bar B^0) \to \rho^0\gamma$ and $B^\pm \to \rho^\pm
\gamma$. Averaged over the charge conjugated modes, the 
current $\cl{90}$ upper limits are~\cite{Jessop02}:
\bea
{\cal B}(B^0 \to \rho^0 \gamma) &<& 1.4 \times 10^{-6} \; , \\
{\cal B}(B^\pm \to \rho^\pm \gamma) &<& 2.3 \times 10^{-6} \; , \\
{\cal B}(B^0 \to \omega \gamma) &<& 1.2 \times 10^{-6} \; .
\eea
They have been combined,
using isospin weights for $B \to \rho \gamma$ decays and assuming
${\cal B}(B^0 \to \omega \gamma)={\cal B}(B^0 \to \rho^0 \gamma)$,
to yield the improved upper limit~\cite{Jessop02}
\beq
{\cal B}(B \to \rho \gamma) < 1.9 \times 10^{-6}\; .
\label{brhobabar}
\eeq
Together with the current measurements of the branching
ratios for $B \to K^* \gamma$ decays by BABAR, this yields
a $\cl{90}$ upper limit~\cite{Jessop02}
\bea
& & R(\rho \gamma/K^*\gamma) \equiv  \frac{{\cal B}(B \to \rho 
\gamma)}{{\cal B}(B\to K^* \gamma)} < 0.047 \; .
\eea
Thus, we see that the current experimental upper limit on the ratio
$R (\rho \gamma /K^* \gamma)$ is typically a factor 2 away from the
central values of the SM. This can be seen graphically in Fig.~1, where
we show the fit of the unitarity triangle resulting from the various
direct and indirect measurements, including
the CP asymmetry $a_{\psi K_s}$, reported by the BABAR and BELLE
collaboration at this conference, with the current world average being
$a_{\psi K_s}=0.734 \pm 0.054$~\cite{Nir:2002gu}. The 
(almost concentric) constraints in the $(\bar{\rho},\bar{\eta})$ plane
from the measured value  $\Delta M_{B_d}=0.503 \pm
0.006~\mbox{ps}^{-1}$ \cite{Stocchi02}, and the two competing bounds 
$\Delta M_{B_s} >~14.4~\mbox{ps}^{-1}$
\cite{Stocchi02}, and $R (\rho \gamma /K^* \gamma) < 0.047$
\cite{Jessop02} are also shown. The two set of curves for the constraints 
following from
$\Delta M_{B_d}$ and $\Delta M_{B_s}$, shown with and without the
chiral logs, reflect the current uncertainty
inherent in Lattice calculations due to chiral extrapolations
from the region $m_s$ to $m_d$~\cite{Lellouch:2002}. We see that the
quantity $R (\rho \gamma /K^* \gamma)$ is not yet competitive with the 
other two constraints from either $\Delta M_{B_d}$ or $\Delta M_{B_s}$. 
However, this sure will change 
as the B factory experiments will soon reach the SM-sensitivity on 
the decay mode $B \to \rho \gamma$, and hence on $R (\rho \gamma /K^* 
\gamma)$. 
\begin{figure}[htb]
\vspace{10pt}
\includegraphics[width=7.5cm]{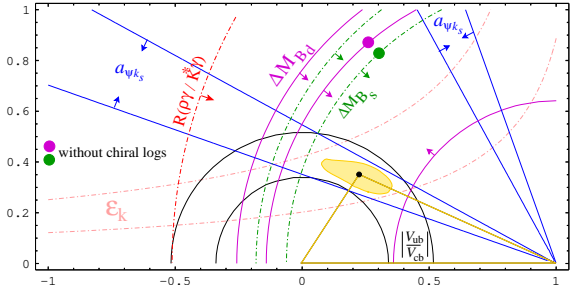}
\vskip 0.2cm
\caption{\it Unitary triangle fit in the SM and
the resulting 95\% C.L. contour in the $\bar \rho$ - $\bar \eta$  
plane. The impact of the $R(\rho\gamma/K^*\gamma) < 0.047$ constraint is
also shown. (From  Ref.~\cite{Ali:2002kw}.)}
\label{fig:utsm}
\vskip -0.2cm
\end{figure}  
\section{THE DECAYS $B \to (X_s,K,K^*)\ell^+ \ell^-$}
To discuss the FCNC semileptonic decays in the SM, one has to extend the 
operator basis given in Eq.~(\ref{Heff}) by adding two semileptonic 
operators:
\begin{eqnarray}
    O_9    & = & \frac{e^2}{g_s^2}(\bar{s}_L\gamma_{\mu} b_L)
                \sum_\ell(\underbrace{\bar{\ell}\gamma^{\mu}\ell}_{\rm
V}) \, , \nonumber\\ 
    O_{10} & = & \frac{e^2}{g_s^2}(\bar{s}_L\gamma_{\mu} b_L)
                \sum_\ell(\underbrace{\bar{\ell}\gamma^{\mu} \gamma_{5}
\ell}_{\rm A}),
\end{eqnarray}
where $e$ ($g_s)$ is the QED (QCD) coupling constant and the subscript $L$ 
refers to the left-handed components of the 
fermion fields. There are additional non-local contributions, which can be 
added to $C_9$, defining an {\it effective} Wilson coefficient,
which is a misnomer as it is actually a function of the dilepton mass 
squared  ($\hat{s}=s/M_B^2$):
$ C_9^{\rm eff}(\hat{s}) = C_9 \eta(\hat{s}) + Y(\hat{s})$, where
$\eta(\hat{s})=(1+ O(\alpha_s))$ includes the real  
gluon bremsstrahlung and virtual contributions to the matrix element of
the operator $O_9$ \cite{Jezabek:1988ja}, and
 $Y(\hat{s})$ contains perturbative charm loops and Charmonium
resonances ($J/\psi, \psi^\prime,...$).

Concentrating on the perturbative contribution only, which is 
expected to be the dominant contribution away from the resonances, and
including the leading power corrections in $1/m_b$ and $1/m_c$, 
the dilepton invariant mass distribution for $B \to X_s \ell^+ \ell^-$ 
can be written as~\cite{Ali:2002jg}:
\begin{eqnarray}
    \frac{d\Gamma(b\to s \ell^+\ell^-)}{d\hat s} &=&
    \left(\frac{\alpha_{em}}{4\pi}\right)^2
    \frac{G_F^2 m_{b,pole}^5\left|\lambda_{ts}\right|^2}
    {48\pi^3} \nn \\
  && \hspace*{-3.0cm} \times (1-\hat s)^2 
 \left[ \left (1+2\hat s\right)
    \left (\left |\widetilde C_9^{\eff}\right |^2+
    \left |\widetilde C_{10}^{\eff}\right |^2 \right ) G_1( \hat s)
\right. \nn \\
    && \hspace*{-3.0cm} \left. + 4\left(1+2/\hat s\right)\left
    |\widetilde C_7^{\eff}\right |^2 G_2( \hat s) +
    12 \mbox{Re}\left (\widetilde C_7^{\eff}
    \widetilde C_9^{\eff*}\right ) \right. \nn \\
 && \hspace{-3.0cm} \left. \times G_3( \hat s) + G_c(\hat s) \right]  \, ,
\label{mfbsll}
\end{eqnarray}
where the functions $G_i( \hat s)$ ($i=1,2,3)$ encode the $1/m_b$ 
corrections~\cite{Ali:1996bm,Falk:1993dh}
and the function $G_c(\hat{s})$ the corresponding $1/m_c$ 
corrections~\cite{Buchalla:1997ky}. Their explicit forms as well as of the
other effective coefficients in the above equation can be seen in 
Ref.~\cite{Ali:2002jg}. As noted in Ref.~\cite{Asatrian:2001de}, inclusion 
of 
the explicit $O(\alpha_s)$ corrections reduces the branching ratio for $B 
\to X_s \ell^+ \ell^-$, compared to the partial $O(\alpha_s)$-corrected 
result~\cite{Ali:1996bm} and its scale dependence is considerably reduced.

Data from the BELLE collaboration on the dilepton
invariant mass spectrum has been analyzed
using Ref.~\cite{Ali:2002jg}, which incorporates all the explicit
$O(\alpha_s)$ and power corrections mentioned above. The hadron invariant
mass spectrum in $B \to X_s \ell^+ \ell^-$, calculated in the heavy quark
effective theory (HQET) and in a phenomenological Fermi motion model
\cite{Ali:1998nq}, has been used to estimate the effect of the   
experimental cuts on $M_{X_S}$. The results are given in Table 1, and 
are to be understood with a cut on the dilepton invariant mass 
$M_{\ell^+\ell^-} > 0.2$ GeV. Given the current experimental errors, the
agreement with the SM-based estimates is reasonably good.

 The corresponding results for the exclusive decays $B \to K \ell^+ 
\ell^-$, averaged over $\ell= e, \mu$, and for $B \to K^* \ell^+ \ell^-$,
with $\ell=e,\mu$, presented at this 
conference~\cite{Nishida02,Richman02} are also 
given in Table 1 and compared with the SM-based estimates  
\cite{Ali:1999mm,Ali:2002jg}. One notes the comparatively larger 
theoretical uncertainty in the exclusive decay rates due to the form 
factors. However, within the theoretical and 
experimental errors, SM accounts well the current data on all the 
inclusive and exclusive rare $B$ decays. This can be 
used to constrain models of physics beyond-the-SM ~\cite{Ali:2002jg}.

\begin{table*}[htb]
\caption{SM predictions $(O(\alpha_s)$ and leading power corrected) and 
comparison with data (in units of $10^{-6}$)}
\label{table:1}
\newcommand{\cc}[1]{\multicolumn{1}{c}{#1}}
\renewcommand{\tabcolsep}{2pc} 
\renewcommand{\arraystretch}{1.2} 
\begin{tabular}{@{}llll}
\hline
Decay Mode & Theory (SM)~\cite{Ali:1999mm,Ali:2002jg} & 
BELLE~\cite{Nishida02} &  
BABAR~\cite{Richman02} \gh \\ \hline
$B\to K\ell^+\ell^-$&$ 0.35\pm 0.12$
&   $0.58^{+0.17}_{-0.15} \pm 0.06 $ &  $0.78^{+0.24
+0.11}_{-0.20 -0.18}$  \gh \\
 \hline
$B\to K^*e^+e^-$ & $1.58\pm 0.52 $ &  $<5.1$ & $1.68 ^{+0.68}_{-0.58} \pm 
0.28$ \gh \\
\hline
$B\to K^*\mu^+\mu^-$ & $1.2\pm 0.4$ & $< 3.0$ & $ < 3.0$;
 weighted $e^+ e^-$,
$\mu^+ \mu^-$ \gh \\ \hline
$B\to X_s \mu^+ \mu^-$ & $4.15 \pm 0.70$ & $7.9 \pm 2.1^{+2.0}_{-1.5}$
 & -- \gh
 \\ \hline
$B\to X_s e^+ e^-$ & $4.2 \pm 0.7$ &  $ 5.0 \pm 2.3 ^{+1.2}_{-1.1}$ &  -- 
\gh \\
\hline
$B\to X_s \ell^+ \ell^-$ & $4.18 \pm 0.70$ &  $ 6.1\pm
1.4^{+1.3}_{-1.1}$ &  -- \gh \\
\hline
\end{tabular}\\[2pt]
The inclusive measurements and the corresponding SM rates are given with a 
cut on the dilepton mass, $M_{\ell^+\ell^-}> 0.2$ GeV.
\end{table*}


\begin{thebibliography}{9}
%
\bibitem{Ammar:1993sh}
R.~Ammar {\it et al.}  (CLEO Collaboration),
Phys.\ Rev.\ Lett.\  {\bf 71}, 674 (1993).

\bibitem{Alam:1995aw}
M.~S.~Alam {\it et al.}  (CLEO Collaboration),
Phys.\ Rev.\ Lett.\  {\bf 74} (1995) 2885.

\bibitem{Chen:2001fj}
S.~Chen {\it et al.}  (CLEO Collaboration),
Phys.\ Rev.\ Lett.\  {\bf 87} (2001) 251807
[hep-ex/0108032].

%
\bibitem{Coan:1999kh}
T.~E.~Coan {\it et al.}  (CLEO Collaboration),
Phys.\ Rev.\ Lett.\  {\bf 84} (2000) 5283
[hep-ex/9912057].
%
%
\bibitem{Barate:1998vz}
R.~Barate {\it et al.}  (ALEPH Collaboration),
Phys.\ Lett.\ B {\bf 429} (1998) 169.

\bibitem{Jessop02}
C. Jessop (BABAR Collaboration), these proceedings.

\bibitem{Nishida02}
S. Nishida (BELLE Collaboration), these proceedings.

%
\bibitem{bdgAP}
A.~Ali and A.~Y.~Parkhomenko, Eur.\ Phys.\ J.\ C {\bf 23} (2002) 89
[hep-ph/0105302].
%
\bibitem{Bosch:2001gv}
S.~W.~Bosch and G.~Buchalla, Nucl.\ Phys.\ B {\bf 621} (2002) 459
[hep-ph/01060].
%
%
\bibitem{Beneke:2001at}
M.~Beneke, T.~Feldmann and D.~Seidel, Nucl.\ Phys.\ B {\bf 612} (2001) 25
[hep-ph/0106067].
%
%
\bibitem{Beneke:1999br}
M.~Beneke, G.~Buchalla, M.~Neubert and C.~T.~Sachrajda,
Phys.\ Rev.\ Lett.\  {\bf 83} (1999) 1914
[hep-ph/9905312].
%
%
%
\bibitem{Ali:2000zu}
A.~Ali, L.~T.~Handoko and D.~London,
Phys.\ Rev.\ D {\bf 63} (2000) 014014
[hep-ph/0006175].
%
%
\bibitem{Ali:2002kw}
A.~Ali and E.~Lunghi,
DESY-02-089, hep-ph/0206242.
%

\bibitem{Richman02}
J. Richman (BABAR Collaboration), these proceedings.

\bibitem{Bobeth:1999mk}
C.~Bobeth, M.~Misiak and J.~Urban,
Nucl.\ Phys.\ B {\bf 574}, 291 (2000)
[hep-ph/9910220].


\bibitem{Asatrian:2001de}
H.~H.~Asatrian, H.~M.~Asatrian, C.~Greub and M.~Walker,
Phys.\ Lett.\ B {\bf 507}, 162 (2001)
[hep-ph/0103087].

\bibitem{Ghinculov:2002pe}
A.~Ghinculov, T.~Hurth, G.~Isidori and Y.~P.~Yao,
[hep-ph/0208088].


\bibitem{Asatrian:2002va}
H.~M.~Asatrian, K.~Bieri, C.~Greub and A.~Hovhannisyan,
[hep-ph/0209006].

\bibitem{Ali:1996bm}   
A.~Ali, G.~Hiller, L.~T.~Handoko and T.~Morozumi,
Phys.\ Rev.\ D {\bf 55}, 4105 (1997)
[hep-ph/9609449].

\bibitem{Falk:1993dh}
A.~F.~Falk, M.~E.~Luke and M.~J.~Savage,
Phys.\ Rev.\ D {\bf 49}, 3367 (1994)
[hep-ph/9308288].

\bibitem{Buchalla:1997ky}
G.~Buchalla, G.~Isidori and S.~J.~Rey,
Nucl.\ Phys.\ B {\bf 511}, 594 (1998)
[hep-ph/9705253].


\bibitem{Ali:1994bf}
A.~Ali, G.~F.~Giudice and T.~Mannel,
Z.\ Phys.\ C {\bf 67}, 417 (1995)
[hep-ph/9408213].

   
\bibitem{Beneke:2000wa}
M.~Beneke and T.~Feldmann,
Nucl.\ Phys.\ B {\bf 592}, 3 (2001)
[hep-ph/0008255].

\bibitem{Charles:1998dr}  
J.~Charles, A.~Le Yaouanc, L.~Oliver, O.~Pene and J.~C.~Raynal,
Phys.\ Rev.\ D {\bf 60}, 014001 (1999)
[hep-ph/9812358].


\bibitem{Ali:2002qc}
A.~Ali and A.~S.~Safir,
[hep-ph/0205254].

\bibitem{Ali:1999mm}
A.~Ali, P.~Ball, L.~T.~Handoko and G.~Hiller,
Phys.\ Rev.\ D {\bf 61}, 074024 (2000)
[hep-ph/9910221].

\bibitem{Ali:2002jg}
A.~Ali, E.~Lunghi, C.~Greub and G.~Hiller,
Phys.\ Rev.\ D {\bf 66}, 034002 (2002)
[hep-ph/0112300].

\bibitem{Gambino:2001ew}
P.~Gambino and M.~Misiak,
Nucl.\ Phys.\ B {\bf 611} (2001) 338
[hep-ph/0104034].
%
\bibitem{Buras:2002tp}
A.J.~Buras, A.~Czarnecki, M.~Misiak and J.~Urban,
Nucl.\ Phys.\ B {\bf 631} (2002) 219
[hep-ph/0203135].
%

\bibitem{Aubert:2002pd}
B.~Aubert {\it et al.}  [BaBar Collaboration],
[hep-ex/0207076].
%
%
%
\bibitem{Abe:2001hk}
K.~Abe {\it et al.}  (BELLE Collaboration),
Phys.\ Lett.\ B {\bf 511} (2001) 151
[hep-ex/0103042].
%

\bibitem{Chetyrkin:1996vx}
K.~G.~Chetyrkin, M.~Misiak and M.~M\"unz,
Phys.\ Lett.\ B {\bf 400}, 206 (1997)
[Erratum-ibid.\ B {\bf 425}, 414 (1998)]
[hep-ph/9612313].

%
\bibitem{Misiak-02}
M. Misiak, talk presented at the DESY-Theory Workshop, Hamburg, September
24-26, 2002, and private communication.

%
\bibitem{Hagiwara:pw}
K.~Hagiwara {\it et al.}  (Particle Data Group Collaboration),
Phys.\ Rev.\ D {\bf 66} (2002) 010001.
%
\bibitem{AM-02}
A. Ali and M. Misiak, in Proceedings of the CERN-CKM Workshop, Geneva, 
Feb.~13-16, 2002.


\bibitem{Soares:1991te}
J.~M.~Soares,
Nucl.\ Phys.\ B {\bf 367}, 575 (1991).

\bibitem{Greub:1994tb}
C.~Greub, H.~Simma and D.~Wyler,
Nucl.\ Phys.\ B {\bf 434}, 39 (1995)
[Erratum-ibid.\ B {\bf 444}, 447 (1995)]
[hep-ph/9406421].


\bibitem{Greub:1996tg}
C.~Greub, T.~Hurth and D.~Wyler,
Phys.\ Rev.\ D {\bf 54}, 3350 (1996)
[hep-ph/9603404].

\bibitem{Ali:1995uy}
A.~Ali and V.~M.~Braun,
Phys.\ Lett.\ B {\bf 359}, 223 (1995)
[hep-ph/9506248].
 
\bibitem{Khodjamirian:1995uc}
A.~Khodjamirian, G.~Stoll and D.~Wyler,
Phys.\ Lett.\ B {\bf 358}, 129 (1995)
[hep-ph/9506242].

\bibitem{Grinstein:2000pc}
B.~Grinstein and D.~Pirjol,
Phys.\ Rev.\ D {\bf 62}, 093002 (2000)
[hep-ph/0002216].

\bibitem{Kagan:2001zk}
A.~L.~Kagan and M.~Neubert,
Phys.\ Lett.\ B {\bf 539}, 227 (2002)
[hep-ph/0110078].
%

\bibitem{Ball:1998kk}
P.~Ball and V.~M.~Braun,
Phys.\ Rev.\ D {\bf 58}, 094016 (1998)
[hep-ph/9805422].


\bibitem{DelDebbio:1997kr}
L.~Del Debbio, J.~M.~Flynn, L.~Lellouch and J.~Nieves  [UKQCD 
Collaboration],
Phys.\ Lett.\ B {\bf 416}, 392 (1998)
[hep-lat/9708008].

\bibitem{Becirevic02}
D. Becirevic, these proceedings.

%
\bibitem{Narison:1994kr}
S.~Narison,
Phys.\ Lett.\ B {\bf 327} (1994) 354
[hep-ph/9403370].
%
\bibitem{Melikhov:2000yu}
D.~Melikhov and B.~Stech,
Phys.\ Rev.\ D {\bf 62} (2000) 014006
[hep-ph/0001113].
%
\bibitem{Nir:2002gu}
Y.~Nir,
hep-ph/0208080.
%
\bibitem{Stocchi02}
A. Stocchi, these proceedings.

\bibitem{Lellouch:2002}
L. Lellouch, these proceedings.
%
\bibitem{Jezabek:1988ja}
M.~Jezabek and J.~H.~K\"uhn,
Nucl.\ Phys.\ B {\bf 320}, 20 (1989).

\bibitem{Ali:1998nq}
A.~Ali and G.~Hiller,
Phys.\ Rev.\ D {\bf 58}, 074001 (1998)
[hep-ph/9803428].


%
\end{thebibliography}
\end{document}